\documentclass[12pt,preprint]{aastex}
\usepackage{epsf}

\newcommand{\be}{\begin{displaymath}}
\newcommand{\ee}{\end{displaymath}}
\newcommand{\bea}{\begin{eqnarray}}
\newcommand{\eea}{\end{eqnarray}}

\shortauthors{Pavel Denissenkov}
\shorttitle{Thermohaline Mixing in Low-Mass Red Giants}

\begin{document}

\title{THERMOHALINE MIXING: DOES IT REALLY GOVERN THE ATMOSPHERIC
       CHEMICAL COMPOSITION OF LOW-MASS RED GIANTS?}

\author{Pavel A. Denissenkov\altaffilmark{1}, and William J. Merryfield\altaffilmark{2}}
\altaffiltext{1}{Department of Physics \& Astronomy, University of Victoria,
       P.O.~Box 3055, Victoria, B.C., V8W~3P6, Canada}
              \email{pavel.denisenkov@gmail.com}
\altaffiltext{2}{Canadian Centre for Climate Modelling and Analysis, University of Victoria,
       P.O.~Box 3065, Victoria, B.C., V8W~3V6, Canada}
              \email{bill.merryfield@ec.gc.ca}
 
\begin{abstract}
First results of our 3D numerical simulations of thermohaline convection driven by $^3$He burning
in a low-mass RGB star at the bump luminosity are presented. They confirm our previous conclusion
that this convection has a mixing rate which is a factor of 50 lower than the observationally
constrained rate of RGB extra-mixing. It is also shown that the large-scale instabilities of
salt-fingering mean field (those of the Boussinesq and advection-diffusion equations averaged over
length and time scales of many salt fingers), which have been observed to increase the rate of oceanic
thermohaline mixing up to one order of magnitude, do not enhance the RGB thermohaline mixing.
We speculate on possible alternative solutions of the problem of RGB extra-mixing, among which
the most promising one that is related to thermohaline mixing is going to take advantage of 
the shifting of salt-finger spectrum towards larger diameters by toroidal magnetic field.

\end{abstract} 


\section{Introduction}

Red giants with $M\la 1.5\,M_\odot$ are known to experience extra-mixing in their convectively stable
radiative zones that separate the hydrogen burning shell (HBS) from the bottom of the convective envelope (BCE).
It operates during their first ascent along the red giant branch (RGB) above the bump luminosity, after    
the HBS has crossed a chemical composition discontinuity left behind by the BCE at the end of the first dredge-up.
This causes the atmospheric $^{12}$C/$^{13}$C ratio and carbon abundance
to resume their declines with an increase in luminosity.

The most promising physical mechanism of RGB extra-mixing is thermohaline convection (\citealt{chz07a}).
It is driven by a double-diffusive instability that occurs when
a scalar component that stabilizes a density stratification, e.g. the temperature $T$, diffuses away faster than a destabilizing component,
e.g. the mean molecular weight $\mu$. \cite{eea06} brought our attention to the fact that the reaction
$^3$He($^3$He,\,2p)$^4$He produces a local depression of $\mu$, $\Delta\mu\sim -10^{-4}$,
in the HBS tail above the bump luminosity. As a result, $\mu$ is increasing with the radius, which
turns the $\mu$ profile into the destabilizing component. Moreover, the radiative thermal diffusivity $K$ exceeds the molecular
diffusivity $\nu_{\rm mol}$ by many orders of magnitude (the inverse Lewis number $\tau = \nu_{\rm mol}/K\sim 10^{-6}$ in the region of $\mu$ depression),
which should lead to the double-diffusive instability.
In the ocean, a similar instability develops where warm salty water overlies cold fresh water. The oceanic thermohaline mixing     
usually takes the form of vertically elongated fluid parcels of rising fresh and sinking salty water
adjacent to each other, called ``salt fingers'' (\citealt{s60}). It has extensively been studied both experimentally and theoretically (e.g., \citealt{k03}).

The efficiency of thermohaline mixing in stellar cases can be estimated only theoretically, e.g. via a linear stability analysis of
its governing equations (\citealt{u72}) or via direct numerical simulations.
Denissenkov (2010, hereafter Paper~I) has recently conducted a comparative study of thermohaline convection in
the oceanic and RGB cases using both a linear stability analysis and 2D numerical simulations. Unfortunately, the linear
theory yields a thermohaline diffusion coefficient that is proportional to the square of
the (unknown) aspect ratio, $a=l/d$, of a finger, where $l$ and $d$ are the finger's length and diameter.
Applications of this diffusion coefficient to model the evolutionary decline of carbon abundance
in low-mass red giants demand that $a > 7$ (Paper~I), which is in agreement with the original result of \cite{chz07a}.
However, our 2D numerical simulations of $^3$He-driven thermohaline convection in the vicinity of a $\mu$ depression
yield a mixing rate equivalent to that approximated by the linear-theory diffusion coefficient with $a < 1$, i.e. it is
a factor of $\sim 1/50^{\rm th}$ of its observationally constrained value. The difference in the finger aspect ratios
between the oceanic ($a > 1$) and RGB ($a < 1$) cases is most likely determined by the very low viscosity $\nu$,
or the Prandtl number $Pr = \nu/K\sim 10^{-6}$, in the RGB case. This facilitates the development of secondary shear instabilities
that strongly limit the growth of salt fingers (\citealt{r10}, and Paper~I). On the other hand, the
diffusion coefficient for thermohaline convection with $a > 7$ has been shown to reproduce many of 
the RGB extra-mixing observational patterns at all metallicities so well
that we had to postpone further discussion of these findings until we were able to confirm or refute
the results of our 2D numerical simulations using 3D computations.

In this Letter, we present first results of our 3D numerical simulations of thermohaline convection for
the oceanic and RGB cases. It turns out that they do not differ much from the 2D results reported in Paper~I. 
We also consider a possibility, mentioned in Paper~I, that
RGB thermohaline mixing may be enhanced by large-scale mean-field instabilities, such as the collective, $\gamma$-, and
intrusive ones, that are known to occur in the ocean. Finally, we speculate on possible alternative solutions of the problem of
RGB extra-mixing.

\section{3D Numerical Simulations of Thermohaline Convection}

The basic equations and numerical techniques that are employed in our 3D simulations of thermohaline convection for
the oceanic and RGB cases are described in detail by \cite{gea03}. They are very similar to those
used in Paper~I. The main goal of Gargett et al. was to study differential scalar diffusion in
3D stratified turbulence. Therefore, they considered a density stratification stable against the double-diffusive
instability and used a specified turbulent velocity field as an initial condition. To apply their method and
computer code for our purposes, we have made the slower diffusing component, which represents salinity
$S$ in the ocean and $\mu$ in the current investigation, destabilizing by changing 
the sign of the vertical velocity term in their equation (16). We have also reduced the amplitude of the initial velocity field,
so that it serves as a mild perturbation to initiate the double-diffusive instability.
Our available computational resources have allowed us to run the code within reasonable time intervals
with a spatial resolution of $320^3$. Simple estimates of the relevant turbulent kinetic
energy dissipation rate and the Kolmogorov and Batchelor length scales show that this resolution is close
to marginal for our runs, however, not fully resolving the slower-diffusing component has been shown
not to have a large effect on the computed fluxes (\citealt{tea10}). 
The 3D simulations have been carried out for the same number of fastest growing fingers 
in the computational domain and parameter sets from Table~1 in Paper~I that were
used in our 2D computations.

The blue and red solid curves in the upper left panel of Fig.~\ref{fig:f1} show transitions of the ratio
$D_S/k_T$ to its equilibrium values in our 2D and 3D simulations of the oceanic thermohaline convection.
Here, $D_S$ and $k_T$ are the turbulent salt and microscopic thermal diffusivities. The 2D simulations were
performed with the resolution $1024^2$ (Paper~I). Our estimated 3D equilibrium value of $D_S/k_T$ exceeds
its corresponding 2D value merely by a factor of 1.5 (compare the blue and red dashed lines). It is in good agreement with the result obtained
using a better resolution by \cite{tea10} that we have interpolated for the value of density ratio 
$R_\rho = \alpha\Delta T/\beta\Delta S = 1.6$ ($\alpha$ and $\beta$ are the coefficients of thermal expansion
and haline contraction) from their Table~1 (star symbol). The upper right panel makes a similar comparison
for the RGB thermohaline convection. In this case, we do not find any difference
between the 2D and 3D equilibrium values of the ratio $D_\mu/K$ (the blue and red dashed lines coincide).
Therefore, we conclude that our results reported in Paper~I seem to be correct.
In particular, RGB thermohaline convection driven by $^3$He burning yields a turbulent mixing rate that is
by a factor of 50 lower than the value required by observations.

\section{Mean-Field Instabilities}

Observations, laboratory experiments, and direct numerical simulations show that small-scale salt-fingering convection in the ocean
can lead to instabilities and the formation of dynamical structures on much larger scales. These include
the so-called ``collective instability'' leading to the generation of internal gravity waves, the intrusive instability
developing when fluid is stratified both vertically and horizontally (in stars, such a situation may result from
the rotational distortion of level surfaces), and the so-called ``$\gamma$-instability'' that produces thermohaline
staircases (e.g., see \citealt{tea10}, and references therein). The latter structures consist of thick well-mixed convective layers
separated by thin salt-fingering interfaces. The staircases are of special interest
to us because they have been observed to enhance vertical mixing in the ocean by up to an order of magnitude (\citealt{schea05}).
Moreover, our preliminary estimates in Paper~I indicate that the turbulent flux ratio
$\gamma = F_T/F_\mu$ is a decreasing function of $R_\rho$ in the RGB case, which
is a sufficient condition for the $\gamma$-instability, according to \cite{r03}.

The large-scale instabilities are studied by averaging the Boussinesq and advection-diffusion equations for $T$ and $S$ ($\mu$)
over length and time scales of many salt fingers. As a result, a system of mean-field equations is obtained in which
the small-scale turbulent velocity is incorporated into the mean heat and salinity fluxes $F_T$ and $F_S$ (\citealt{tea10}).
The fluxes are related to one another and to the large-scale temperature gradient via the
flux ratio $\gamma = F_T/F_S$ and the Nusselt number $Nu$, which are assumed to depend only on the local value of
the density ratio $R_\rho$; specification of these dependences closes the mean-field equations.
Large-scale changes in $R_\rho$ thus result in modulations of $F_T$ and $F_S$ via the dependences of $\gamma$ and $Nu$
on $R_\rho$ that can either amplify or damp initial perturbations.

\begin{figure}
\epsffile [60 190 480 695] {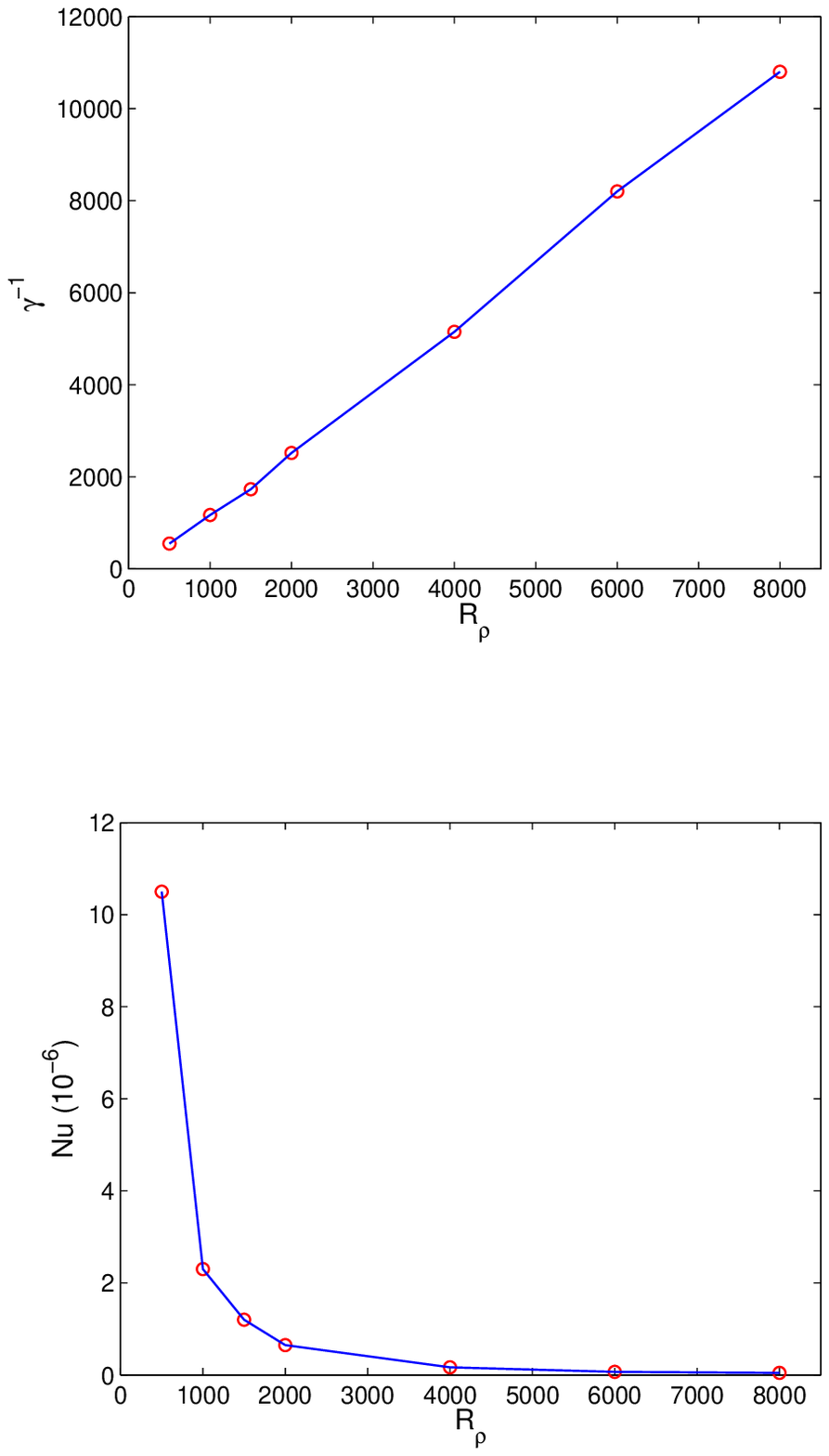}
\caption{The inverse flux ratio, $\gamma^{-1} = F_\mu/F_T$, and Nusselt number as functions of density ratio
         estimated from our 2D numerical simulations of the RGB thermohaline convection.
         }
\label{fig:f2}
\end{figure}

A linear stability analysis of the mean-field equations yields a third-order dispersion relationship
for the growth rates of all the three large-scale instabilities mentioned above (equation 2.13 of \citealt{tea10}).
Coefficients of the cubic equation depend on the Prandtl and inverse Lewis numbers as well as on
$Nu$, $\gamma$ and their derivatives with respect to $R_\rho$. The coefficients are also functions of
the horizontal and vertical wavenumbers, $l$ and $k$. To calculate them, we first need to estimate
the dependences of $Nu$ and $\gamma$ on $R_\rho$. \cite{tea10} estimated those dependences via a body of
3D numerical simulations of the oceanic salt-fingering convection. We have used data from their Table~2
to reproduce, for a comparison, their ``flower plot'' depicting the growth rates of fastest growing
instabilities at $R_\rho = 1.5$ in the absence of lateral gradients (the lower left panel of Fig.~\ref{fig:f1}). 
The bulb of the flower corresponds to the salt-fingering mode that turns out to be a solution of
the general dispersion equation as well. The leaves of the flower represent the gravity waves of the collective instability, whereas
the $\gamma$-instability modes are the ones dominating in the region beneath the leaves.
To obtain estimates of $\gamma(R_\rho)$ and $Nu(R_\rho)$ for the RGB case (Fig.~\ref{fig:f2}), we have performed
additional 2D numerical simulations of the $^3$He-driven thermohaline convection for the same parameter set that was used
in Paper~I but for a number of different values of $R_\rho$. The resulting flower plot is shown in the lower right
panel of Fig.~\ref{fig:f1}. We see that in the RGB case only the salt-fingering mode is unstable in spite of
the flux ratio $\gamma$ being a decreasing function of $R_\rho$, as in the oceanic case.
This difference is explained by the fact that the Nusselt number is extremely small, $Nu\sim 10^{-6}$, in the RGB case.
Because of this, it is necessary to include horizontal diffusive fluxes in the mean-field stability analysis --- something which was
not done by \cite{r03} but was treated by \cite{tea10}. In the RGB case, the inclusion of reasonable lateral gradients
proportional to their corresponding vertical gradients multiplied by the ratio of centrifugal to gravitational acceleration 
only produces an asymmetry of the bulb with respect to the vertical axis.

\section{Possible Alternative Solutions}

\begin{figure}
\epsffile [60 260 580 695] {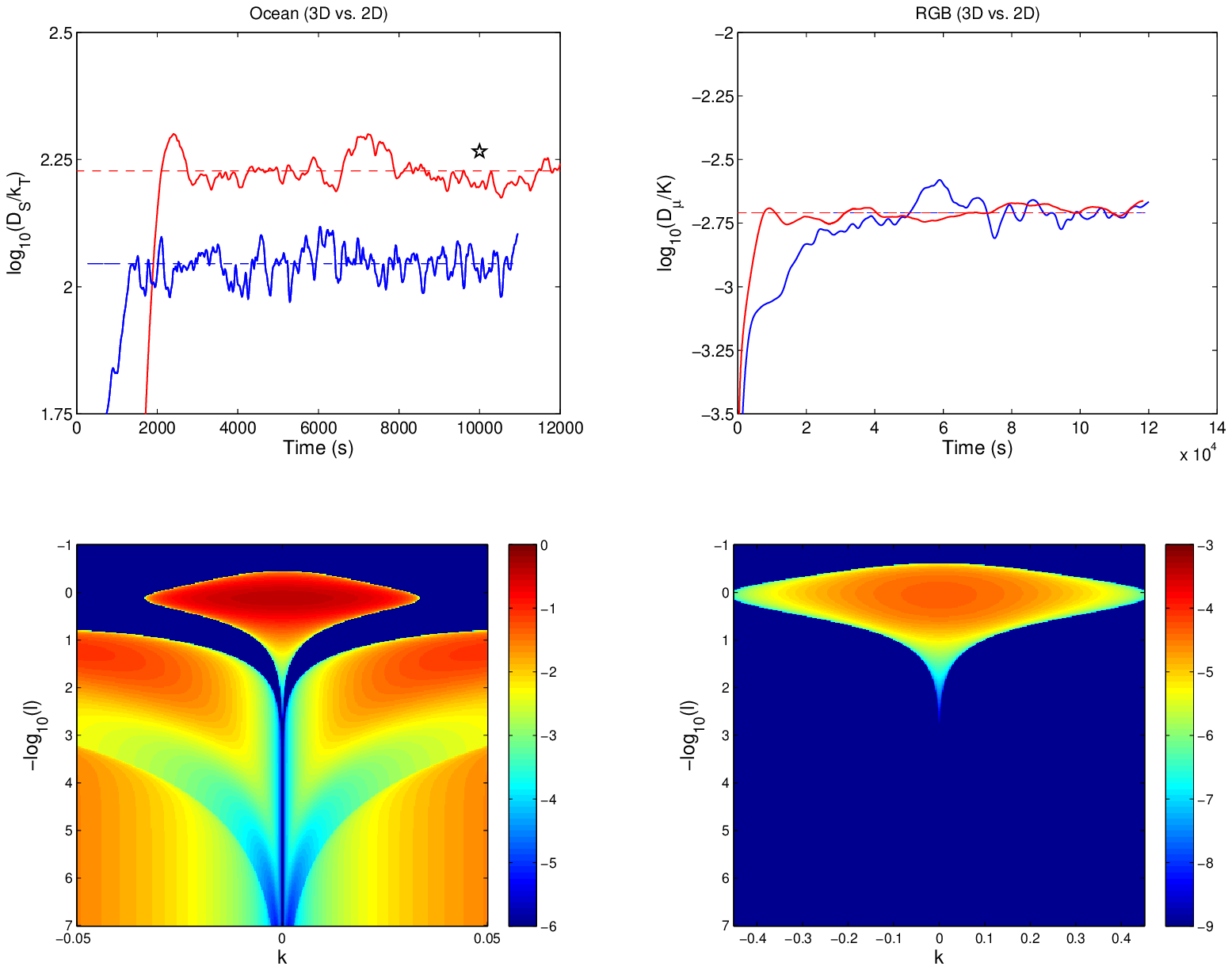}
\caption{Upper panels compare the ratios of thermohaline diffusion coefficient to thermal diffusivity obtained in
         our 3D (red curves) and 2D (blue curves) numerical simulations for the oceanic (left)
         and RGB (right) cases. Black star symbol shows a value interpolated for the density ratio
         $R_\rho = 1.6$ using the higher resolution 3D data of \cite{tea10}. Lower panels depict the decimal logarithms of
         growth rates (s$^{-1}$) of the fastest growing large-scale instabilities of salt-fingering mean-field
         as functions of the horizontal $l$ and vertical $k$ wavenumbers, both measured in cm$^{-1}$, for the oceanic
         (left) and RGB (right) cases.
         }
\label{fig:f1}
\end{figure}

Our 2D and 3D numerical simulations of thermohaline convection driven by $^3$He burning have demonstrated that
it is unlikely to be the sole mechanism of RGB extra-mixing. Its corresponding mixing rate in the vicinity of a $\mu$ depression,
$D_\mu\approx 2\times 10^{-3}\,K$ (the upper right panel of Fig.~\ref{fig:f1}), is too low   
compared to the observationally constrained value of $(D_\mu)_{\rm obs}\approx 0.1\,K$
(Paper~I). Therefore, for its efficiency to be consistent with observations, the RGB thermohaline mixing
has to be either assisted or modified by other processes.
In Paper~I, we have shown that a turbulent viscosity $\nu_{\rm t}$ exceeding the microscopic one by a factor of
$10^4$ could enhance the rate of $^3$He-driven thermohaline mixing up to the observed value, provided that
it was not itself associated with strong turbulent mixing. The enhancement of thermohaline mixing in this case is caused by the fact that
the higher viscosity suppresses the development of secondary shear instabilities that limit the growth of salt
fingers. On the other hand, if the increase of viscosity is accompanied by a similar increase of the rate of turbulent
mixing, the latter will reduce salt-finger buoyancy and vertical transports by smoothing out the chemical composition contrast
between rising and sinking fluid parcels. 

The additional source of turbulence in the radiative zones of
RGB stars could arise from the differential rotation that is produced by the mass inflow that feeds the HBS (Paper~I).
We have employed the COMSOL Multiphysics software package to solve the angular momentum transport equation
for the radiative zone of the low-metallicity bump-luminosity RGB model from Paper~I 
in the presence of mass inflow and turbulent shear mixing. Their respective rate $\dot{r}$ and diffusion coefficient $\nu_{\rm v}$ are given
by equation (23) from Paper~I and equation (3) with the uncertainty factor $f_{\rm v}$ from the paper of \cite{dea06}.
The angular velocity at the BCE was assumed to have the same value, $\Omega_{\rm BCE} = 6.3\times 10^{-7}$ rad\,s$^{-1}$,
as in the M2 bump-luminosity RGB model of \cite{pea06}. Simple estimates show that for $f_{\rm v}\ga 1$ the transport of angular momentum by 
the rotation-induced turbulent diffusion dominates over its advection by meridional circulation, therefore, and for the sake of
simplicity, the latter was neglected. Our computations show that the competition between the terms proportional to $\dot{r}$ and $\nu_{\rm v}$
leads to a stationary solution with $\nu_{\rm v}\approx 3\times 10^{-4}\,K$ in the vicinity of a $\mu$ depression that very weakly depends
on both $f_{\rm v}$ and $\Omega_{\rm BCE}$, although the final rotational shear is a decreasing function of $f_{\rm v}$.
This value exceeds the microscopic viscosity by a factor of $10^2$, which is not enough
to enhance $D_\mu$ to its observed value (Paper~I). It is also difficult to understand why the rotation-induced small-scale turbulence
should not produce a mixing rate of the same magnitude as $\nu_{\rm v}$, in which case $D_\mu$ will not be enhanced at all (Paper~I). 
Therefore, we do not think that the proposed viscosity enhancement is a possible solution of the problem.

Our angular momentum transport computations confirm the conclusion that radiative zones of low-mass RGB stars above the bump
luminosity should possess strong differential rotation. This may lead to the generation of high-amplitude toroidal
magnetic fields, especially close to the HBS, provided that a weak poloidal field is present in the radiative zone. Under these
circumstances, a possible alternative solution of the problem could be magneto-thermohaline mixing (\citealt{dea09}). We intend
to investigate this possibility using 2D numerical simulations in one of our forthcoming papers.

\begin{figure}
\epsffile [60 190 480 695] {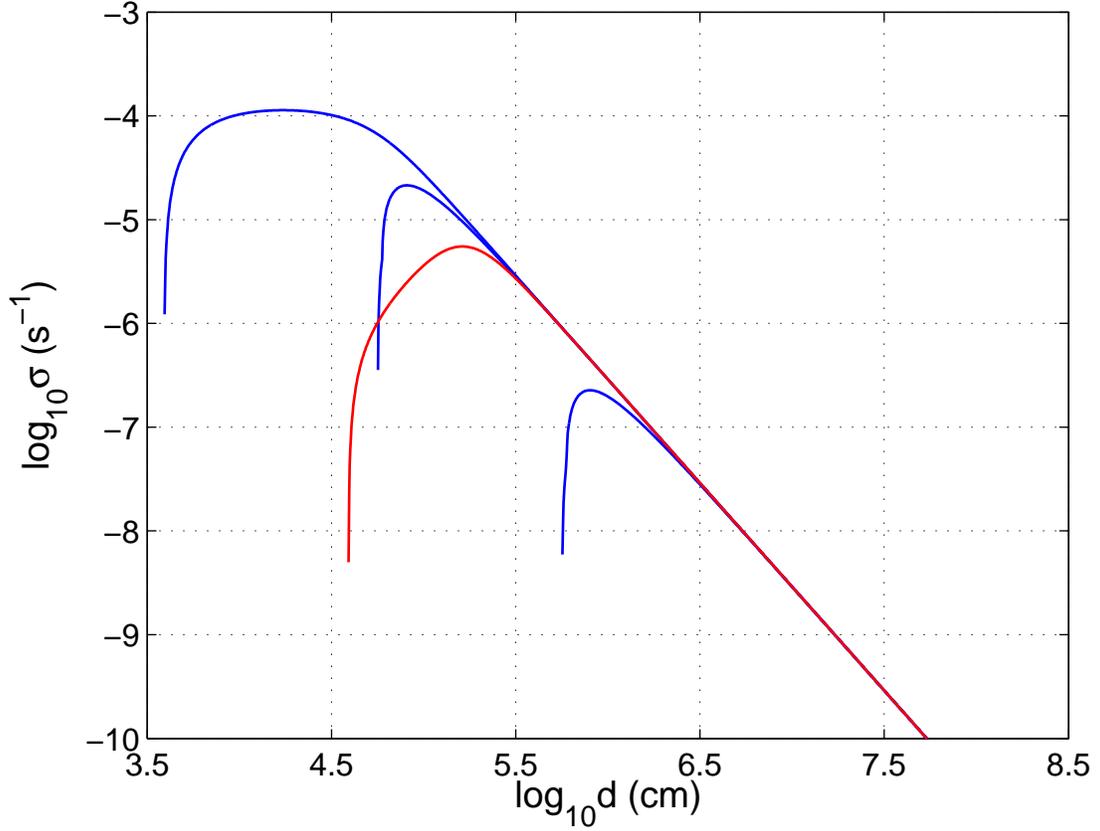}
\caption{The RGB salt-finger growth-rate spectra computed for
         the same parameter set as in Table~1 from Paper~I under the assumptions that
         the toroidal magnetic field has the strengths $B_0=0$, 10, and 100 G (the blue curves from left to right),
         and for $B_0 = 0$ G but with the viscosity amplified by the factor of $10^4$ (the red curve).
         }
\label{fig:f3}
\end{figure}

Toroidal magnetic fields in radiative zones may also
filter out salt fingers with small diameters, so that the maximum growth rate is shifted towards
thicker fingers. Since the finger growth rate decreases with an increase in diameter
as $\sigma\propto d^{-2}$ (e.g., Paper~I), the square of
the vertical velocity shear between the axes of neighbouring rising and sinking fingers, $\sim (\sigma l/d)^2$, is proportional to $a^2 d^{-4}$.
This parameter is likely to control the development of secondary shear instabilities (\citealt{k87}); consequently, thicker fingers
will probably reach higher aspect ratios before they are destroyed.
Following \cite{chz07b}, we have complemented the dispersion
relationship (13) from Paper~I (note that the term $\nu k_Tk_Sk^6$ was erroneously omitted) with terms that take into account
the Lorentz force associated with a specified toroidal magnetic field $B_0$. Its solutions
are shown in Fig.~\ref{fig:f3} for the field strengths $B_0 = 0$, 10, and 100 G (the solid blue curves from left to right). In the case of $B_0=100$ G,
the fastest growing fingers transverse to the field have a diameter that is nearly 40 times as large as that of fingers in the absence of a magnetic field.
This diameter is also almost four times larger than the diameter of fastest growing fingers in the case of a viscosity that is
amplified by the factor of $10^4$ (the red solid curve). However the influence of such a field on the planform of
the fastest growing fingers still needs to be studied. In our next study, we will perform 2D numerical simulations of
the RGB thermohaline convection in the presence of a toroidal magnetic field to find out if the magnetic shift of the salt-finger growth-rate spectrum
towards larger diameters can really enhance its associated mixing rate up to the empirically constrained value.

\acknowledgements
PAD is grateful to Don VandenBerg who has supported this work through his Discovery Grant
from Natural Sciences and Engineering Research Council of Canada.


\end{document}